# Modelling and design of lanthanide ion doped chalcogenide fiber lasers: progress towards the practical realization of the first MIR chalcogenide fiber laser

**Slawomir Sujecki**

1. Wroclaw University of Science and Technology, Wroclaw, Poland; slawomir.sujecki@pwr.edu.pl
* Correspondence: slawomir.sujecki@pwr.edu.pl; Tel.: +48-71-3204588

**Abstract:** This paper presents the progress in the fields of the modelling and design of lanthanide ion doped chalcogenide glass fiber lasers. It presents the laser cavity designs that have been developed in order to optimize the performance of lanthanide ion doped chalcogenide glass fiber lasers. Also various numerical algorithms that have been applied for the optimization of chalcogenide glass lasers are reviewed and compared. The comparison shows that a combination of less accurate but more robust algorithms with more accurate ones gives the most promising performance.

## 1. Introduction

Mid-infrared (MIR) light sources covering spectral range from 3 μm to 25 μm have many potential applications in the fields of biology, medicine, monitoring, agriculture and security. This is because many molecular bonds have resonant oscillation frequencies, which are within MIR wavelength range.

The key enabling technology for exploiting the potential benefits of MIR light is an availability of low cost, robust light source. There are several types of light sources currently available, which cover MIR wavelength range [1]. Wide band light sources include synchrotron, Globar, light emitting diodes (LEDs) and supercontinuum (SC) sources. Synchrotron facilities are available only in several countries in the World. Such facilities are extremely expensive to develop and maintain. Globar on the other hand provides relatively large total output power at a relatively low cost. However, the spectral power density is low, the filament operates at a very high temperature but only a small part of the emitted power can be effectively collected by external optics and used in applications. LEDs on the other hand have very low output power and cover only a small fraction of the entire MIR spectrum. Finally, supercontinuum sources provide a wide wavelength band. Recently SC sources covering wavelengths up to 13 μm were demonstrated [2]. However, MIR SC sources require short pulse, high power seed pump, which is usually realized using either optical parametric oscillators (OPOs) or fiber pulsed lasers. This makes MIR SC sources pricey and limits their robustness. Also the fact that the SC is seeded by a train of pulses limits the range of possible application. There are also narrowband sources that cover MIR wavelength range. These include CO and $CO_2$ lasers, Difference Frequency Generation (DFG), optical parametric oscillators (OPOs) and quantum cascade lasers (QCLs). These sources cover only selected wavelengths. However, some of them can be used to measure wide band MIR spectra by tuning the operating wavelength. This particularly applies to OPOs and QCLs. However, both OPOs and tunable QCLs are very expensive. Also fiber lasers have been demonstrated for MIR wavelength range. Fiber lasers have many characteristics, which make them desirable for a number of specific applications. With fiber lasers it is relatively easy to obtain a single transverse mode operation. Further, fiber lasers facilitate greatly beam delivery. The wide gain of lanthanide ion doped fiber lasers make them desirable when wavelength tune-ability is needed. Also wide gain spectra are useful for generation of ultrashort pulses within the modelocked regime. The so far practically realized fiber lasers for MIR wavelength range use ZBLAN fibers doped with erbium, holmium or dysprosium. Using this technology the longest operating wavelength achieved so far is 3.9 µm [3] when cooling with liquid nitrogen and 3.68 µm [4] in room temperature. Unfortunately, the realization of ZBLAN fiber lasers operating at a wavelength longer than 4 µm has been impeded by the depopulation of higher energetic states via multiphonon transitions. Hence, for the realization of fiber lasers with longer operating wavelengths novel host glasses are needed, which have lower phonon energy than ZBLAN glasses. One of the currently most researched candidates for the realization of this task is the chalcogenide glass.

Chalcogenide glass fibers doped with lanthanide ions have been developed for at least two decades now and show consistently good photoluminescence properties within the MIR wavelength range up to 5.5 µm [5-13]. The observed photoluminescence lifetimes are near to the radiative lifetimes predicted by the Judd-Ofelt theory. The fiber attenuation has been driven down to the dB/km level [14,15]. Lanthanide ion concentrations up to 2000 ppm without crystallization were achieved [6].

A significant progress has also been achieved in developing the technological basis for MIR photonics using the chalcogenide glass fiber in a similar way as it was achieved within the visible and near infrared wavelength range. So far successful splicing with silica fiber has been demonstrated [16]. Chalcogenide glass fiber devices have been demonstrated that allow for beam combining of QCLs [17] and imaging for MIR [18]. Optical attenuators have been demonstrated [19]. Raman lasers have been successfully demonstrated [20]. Also novel technology has been developed to reduce the fiber end facet reflection using direct stamping for nanoimprinting moth-eye structures [21]. Several sensors have been successfully realized using MIR light and chalcogenide glass technology [22-24]. All these achievements show the enormous potential of chalcogenide glass fiber technology, which will hopefully one day lead to a successful realization of lanthanide ion doped chalcogenide glass fiber based lasers with operating wavelengths beyond 4 µm.

This paper provides a review of the progress that has been made so far towards the realization of the first practically working lanthanide ion fiber laser using chalcogenide glass technology in the scientific fields of modelling and design. In the first section a review of the chalcogenide glass fiber technology is given. In the second section laser cavity designs are discussed that have been so far proposed in the available literature for the realization of chalcogenide glass fiber lasers. In the third section the details of the modelling methods that were applied to the design of the chalcogenide glass fiber lasers and to predict their performance is presented. In the last section a summary of the presented review is provided and a discussion of the prospects of the successful realization of the first lanthanide ion doped chalcogenide glass MIR fiber laser is given.

## 2. MIR Fiber Laser Cavity Design

A number of lanthanide ions has been researched on in the context of fiber lasers operating within the MIR wavelength range. A suitable review can be found in [25]. In the context of the chalcogenide glass fiber lasers the ions that have received most attention are: praseodymium, dysprosium and terbium [13,26]. Praseodymium doped into a chalcogenide selenide glass can be pumped using 1480 nm laser diodes, which is of a particular advantage since these laser diodes are commercially available at a relatively low cost when compared with the output power. The reason for this is the fact that 1480 nm laser diodes are applied to pump the erbium doped fiber amplifiers used in optical telecoms. The other pumping wavelengths available for praseodymium are 1550 nm, 2000 nm and the resonant pumping at approximately 4500 nm (Fig.1). For all these wavelengths laser diodes are available. 2000 nm is also suitable for the application of thulium doped fiber lasers. It is also worth noting that at 4500 nm the QCL lasers have the largest available output power, which is an additional advantage. There are energetic levels that allow pumping of praseodymium at shorter wavelengths. However, chalcogenide selenide glass absorption makes pumping at these wavelengths inefficient. In terms of the MIR output wavelengths praseodymium has got two bands that show in the photoluminescence. One of them is around 3700 nm while the other one around 4900 nm with the longest photoluminescence wavelength observe of approximately 5500 nm.

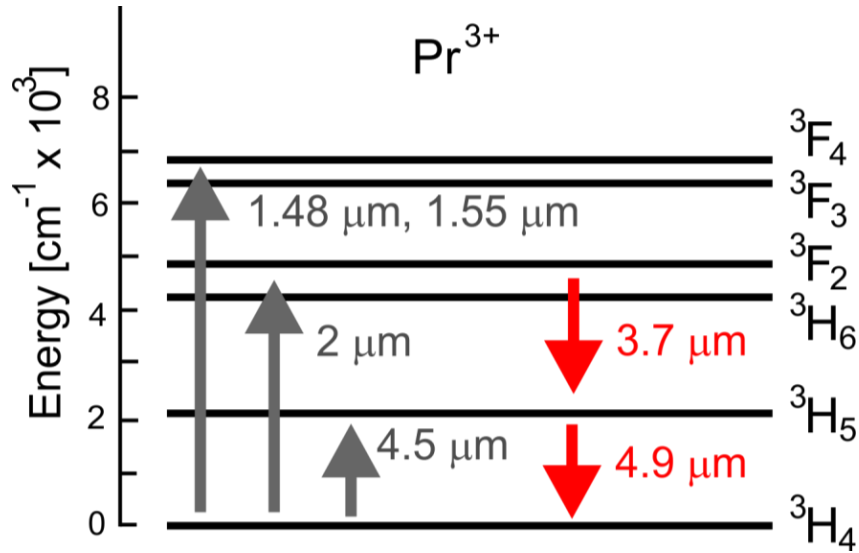

**Figure 1.** Energy level diagram for a praseodymium ion doped into selenide chalcogenide glass

The pumping wavelength available for dysprosium are approximately equal to 1100 nm, 1300 nm, 1700 nm and 2900 nm (Fig.2). Laser diodes with wavelengths corresponding approximately to the first 3 wavelengths are commercially available. 2900 nm can be obtained from either erbium or dysprosium doped ZBLAN fiber laser, which is slightly complicating potential laser realization. In terms of the MIR output spectrum again two bands were experimentally observed in chalcogenide selenide glass: one around 3200 nm and the other one at around 4600 nm with the longest wavelength observed when doped into the chalcogenide selenide glass of approximately 5000 nm. Again, similarly as in the case of praseodymium there are energetic levels that allow pumping of dysprosium at shorter wavelengths but chalcogenide selenide glass absorption makes pumping at these wavelengths inefficient.

For terbium the available pumping wavelengths are approximately 1900 nm, 2000 nm, 2300 nm, 3000 nm and 4500 nm (Fig.3). When compared with dysprosium and praseodymium the commercial availability of laser diodes for pumping terbium is limited. Again 4500 nm is accessed well by high power QCL lasers. 2000 nm is well suited for the application of the thulium fiber lasers. 3000 nm is well suited for holmium doped ZBLAN fiber lasers. There are also laser diodes available for 1900 nm, 2000 nm and 2300 nm but the range of available output powers is limited. For terbium there are no other energetic levels available for pumping in the fundamental configuration. There is also one MIR output band available when doping a chalcogenide selenide glass, which is centered around 5100 nm and stretches up to approximately 5500 nm.

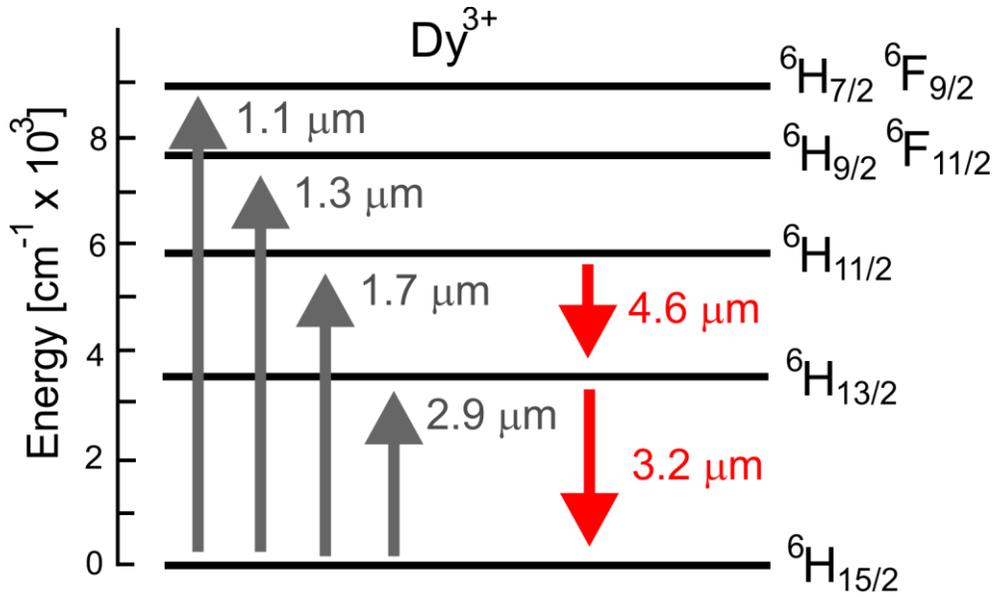

**Figure 2.** Energy level diagram for a dysprosium ion doped into selenide chalcogenide glass

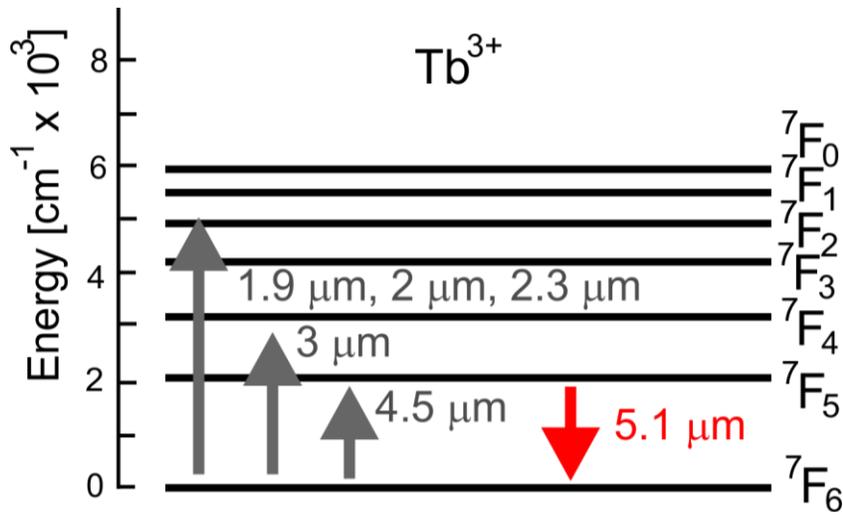

**Figure 3.** Energy level diagram for a terbium ion doped into selenide chalcogenide glass

The first cavity proposed in the literature for a realization of an efficient chalcogenide glass fiber laser was based on dysprosium ions and used the concept of cascade lasing [27]. The schematic diagram of this cavity is shown in Fig.4a. Fiber Grating 1 (FG1) is inscribed in order to trap the idler at around 3100 nm while FG2 forms the cavity for the output signal operating approximately at 4500 nm. The role of the idler is to deplete the lower lasing level $^6H_{13/2}$ by effectively transferring the ions into the $^6H_{15/2}$ level. The signal interacts with the levels $^6H_{11/2}$ and $^6H_{13/2}$.

If the idler wave is not present within the cavity (Fig.4a) then a saturation of the output signal is observed at a level of several mW. Fig. 5 shows a typical dependence of the MIR output signal power on the pump power for a dysprosium doped chalcogenide glass fiber laser operating within the cascade scheme with idler interacting with levels $^6H_{13/2}$ and $^6H_{15/2}$ while signal with $^6H_{11/2}$ and $^6H_{13/2}$ when pumped using approximately 1700 nm. Two regions are distinguishable within the entire range of the pump powers. In the first region the idler power is equal approximately to zero while the signal power initially grows linearly and then saturates. The saturation is due to a long lifetime of the level $^6H_{13/2}$. In the region 2 the idler starts lasing action between the fiber gratings and also to deplete the level $^6H_{13/2}$. Thus idler facilitates the achievement of the population inversion for the signal. In the dependence of the signal power on the pump power this shows as a restoration of a linear growth of the signal power

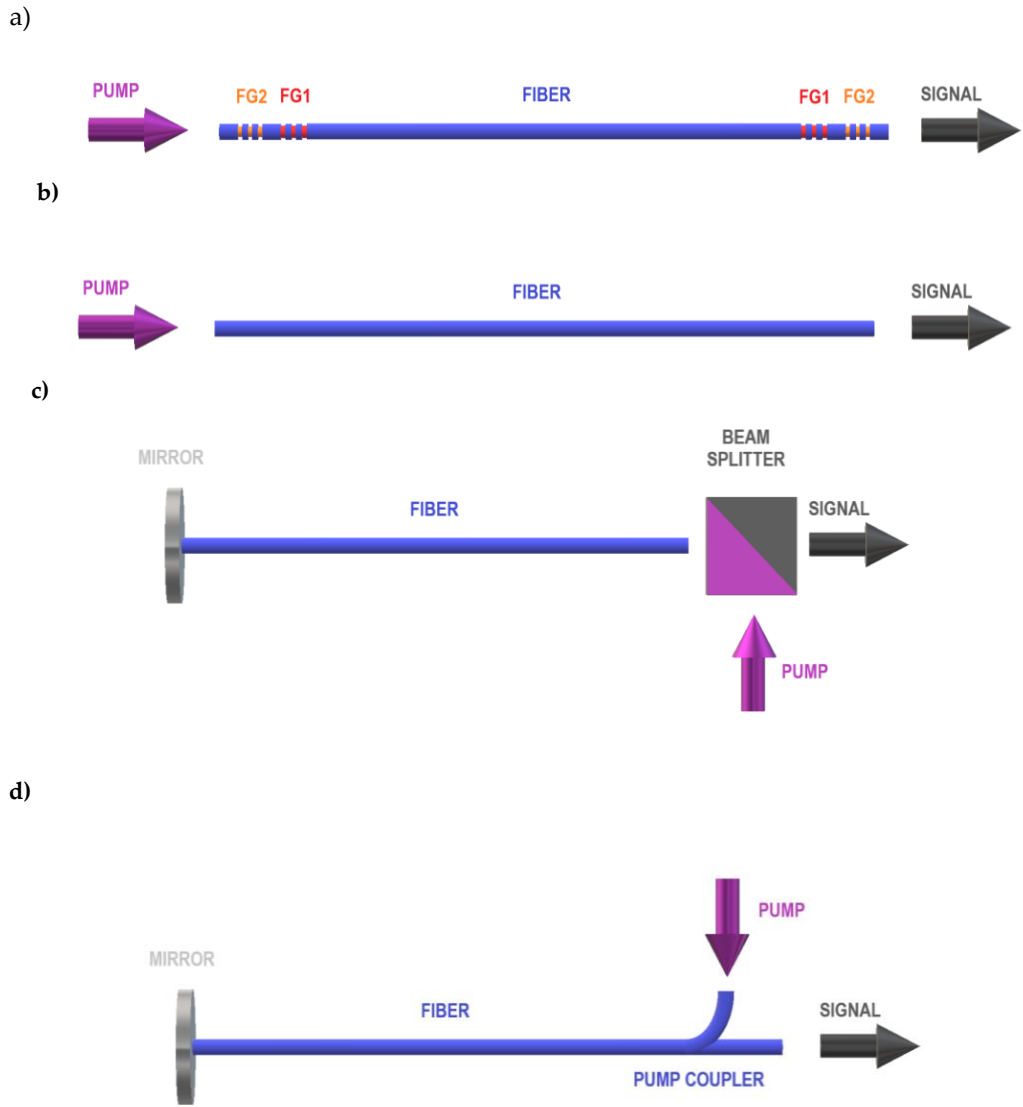

**Figure 4.** Schematic diagrams of cascade lasers (a) with fiber Bragg gratings FG1 and FG2; (b) consisting of the fiber only; (c) a structure using bulk optical elements and (d) all fiber structure using a pump coupler.

The disadvantage of the laser cavity design shown in figure 4a is that its realization is predicated upon a successful fabrication of two pairs of fiber gratings, which is a technological challenge. In [28] therefore some alternative solutions were proposed, which would be simpler to practical implementation. The design shown in Fig.4b consists of a fiber stretch only. This cavity design relies on the large refractive index of chalcogenide glass. As a result a large value of Fresnel reflection from the fiber end facet can be expected. The experimental data suggest that approximately 20 % of Fresnel reflection is achievable. The results obtained show that the efficiency of the cavity design from Fig.4b is much lower than that of Fig.4a. The efficiency can however, be improved by terminating one end of the fiber with a mirror and using a beam splitter at the other end to separate the pump wave from the signal wave (Fig.4c). Alternatively, a pump coupler could be used to make an all fiber structure (Fig.4d). Both the designs from Fig.4c and Fig.4d have similar efficiency at the design from Fig.4a [28]. Hence, they both present a viable alternative to a necessity of fabricating two pairs of fiber gratings.

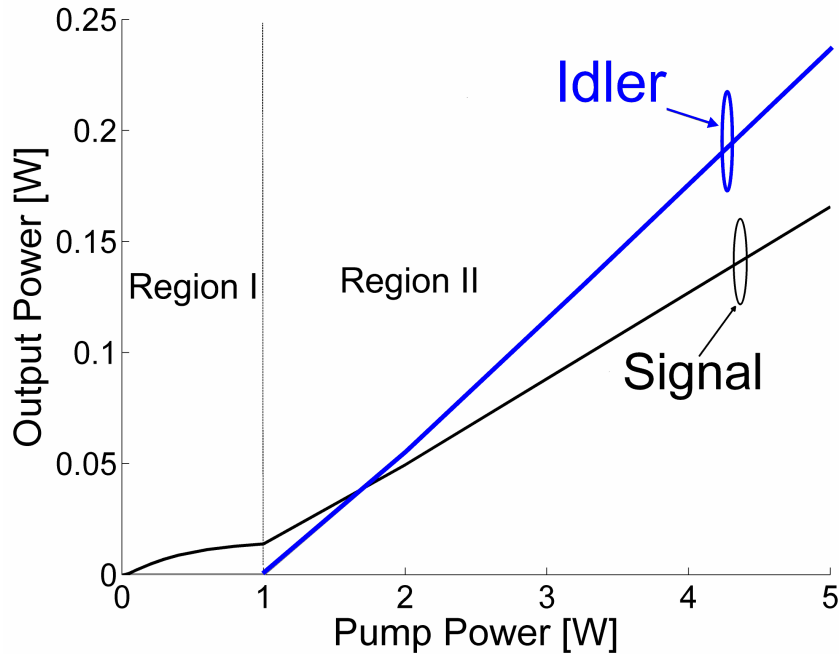

**Figure 5.** An example dependence of the signal and idler power on the pump power for a cascade chalcogenide glass fiber laser doped with dysprosium ions. Figure was adopted from [28].

The cascade lasing scheme was further explored in [26] to study comparatively the dependence of the output power on the pump power for dysprosium, praseodymium and terbium ion doping. The terbium ion doped laser was pumped at 2950 nm whilst the praseodymium ion doped one at 2040 nm. This study showed that the best lasing efficiency can be obtained from a praseodymium ion doped chalcogenide glass fiber laser. This is mainly attributed to a large value of the pump absorption cross section at around 2000 nm when compared with two other ions. The maximum efficiency obtained in this study did not exceed 20 %.

The more recent studies concentrated on the improvement of the lasing efficiency whilst finding alternatives to the cascade pumping scheme [29-32]. In [31] a use of dual pump was proposed to bring the lasing efficiency over 20 %. The dysprosium ion doping was applied and the pumps with wavelengths of 2850 nm and 4092 nm were used. A photonic crystal fiber was proposed to enforce single mode operation and pump and signal wavelengths. A practical realization of a laser cavity with two pump lasers may prove challenging therefore an alternative solution was presented in [32] which allows to improve the laser efficiency with much less challenge with respect to practical realization. By combining a master oscillator with a power amplifier (Fig.6) and using 1700 nm pump only a slope efficiency as high as 38 % can be achieved. The main role of a pump amplifier is to utilize the unused pump power from the master oscillator stage.

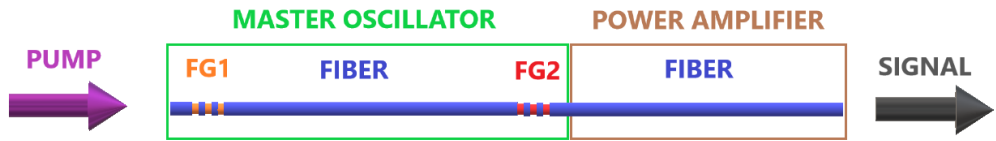

**Figure 6.** Schematic diagram of a master oscillator power amplifier chalcogenide glass fiber laser proposed in [32].

The most recent design that has been proposed is shown in Fig.7. It relies on terbium ion doping and on the fact that terbium ion when doped into chalcogenide selenide glass and pumped at 2950 nm forms a classical 3 level lasing system. This results in a very simple structure of the laser (Fig.7). Due to multiphonon depopulation other pumping wavelengths than the ones using the level $^7F_4$ can be applied also but at the cost of reduced efficiency. The results shown in [10] predict a slope efficiency of more than 40 % despite of a very simple cavity design.

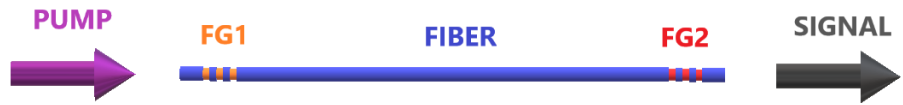

**Figure 7.** Schematic diagram of a chalcogenide glass fiber laser based on a three level scheme realized using terbium ion doping proposed in [10].

In the next section the numerical models used for the design of chalcogenide glass fiber lasers are discussed.

## 3. Numerical Modelling Methods

In literature, which concerns lanthanide ion doped chalcogenide glass fiber modelling and design, the steady state analysis is dominant. It relies on solving the rate equations for the relevant levels with a set of ordinary differential equations that describe the distribution of the pump and signal powers within the cavity. The equations are solved consistently for given boundary conditions that are imposed at the cavity end facets. In order to discuss in more detail the numerical methods used in analysis of chalcogenide glass fiber lasers we focus our discussion on a dysprosium doped chalcogenide glass fiber laser that operates within the cascade lasing scheme pumped at 1710 nm, which was discussed in section 2 (Fig.4, Fig.8) [28]. The algorithms used in this case are identical to those used for the analysis of standard fiber laser cavities and hence the discussed models can be readily applied to the design and analysis of other laser cavities discussed in section 2. For dysprosium ion doped into a chalcogenide selenide glass the rate equations can have a matrix form:

$$\begin{bmatrix} a_{11} & a_{12} & a_{13} \\ a_{21} & a_{22} & a_{23} \\ 1 & 1 & 1 \end{bmatrix} * \begin{bmatrix} N_1 \\ N_2 \\ N_3 \end{bmatrix} = \begin{bmatrix} 0 \\ 0 \\ N \end{bmatrix}, \qquad (1)$$

where N is the total dysprosium concentration and the matrix elements are:

$$a_{11} = \sigma_{pa}\phi_p; \quad a_{12} = \sigma_{\lambda 1a}\phi(\lambda_1); \quad a_{13} = -\sigma_{pe}\phi_p - \sigma_{\lambda 1e}\phi(\lambda_1)\sigma_{\lambda 1a}\phi(\lambda_1) - \frac{1}{\tau_3};$$
$$a_{21} = \sigma_{\lambda 2a}\phi(\lambda_2); \quad a_{22} = -\sigma_{\lambda 2e}\phi(\lambda_2) - \sigma_{\lambda 1a}\phi(\lambda_1) - \frac{1}{\tau_2}; a_{23} = \sigma_{\lambda 1e}\phi(\lambda_1) + \frac{\beta_{32}}{\tau_3};$$

(2)

The indexes used in equations (1) and (2) are easier to follow if one associates levels 1,2 and 3 with levels $^6H_{15/2}$, $^6H_{13/2}$ and $^6H_{11/2}$ from Fig.2, respectively. In equations (1) and (2) $\tau_3$ and $\tau_2$ are the lifetimes of levels $^6H_{11/2}$ and $^6H_{13/2}$, respectively, $\beta_{32}$ is the branching ratio for the $^6H_{11/2}$ -> $^6H_{13/2}$ transition and $\sigma_{xya/e}$ denotes the absorption/emission cross-section for the transition xy. Symbols $\phi_P$, $\phi(\lambda_1)$ and $\phi(\lambda_2)$ denote the photon flux for the pump, signal and idler, respectively. The photon flux $\phi$ is related to optical powers P by the following expressions: $\phi_P = P_P \Gamma_P \lambda_P/(A\,h\,c)$, $\phi(\lambda_1) = P(\lambda_1) \Gamma_{\lambda 1} \lambda_1/(A\,h\,c)$ and $\phi(\lambda_2) = P(\lambda_2) \Gamma_{\lambda 2} \lambda_2/(A\,h\,c)$. A denotes the doping cross-section, h is Planck's constant, while c is the speed of light in free space and the confinement factor is denoted as $\Gamma_x$.

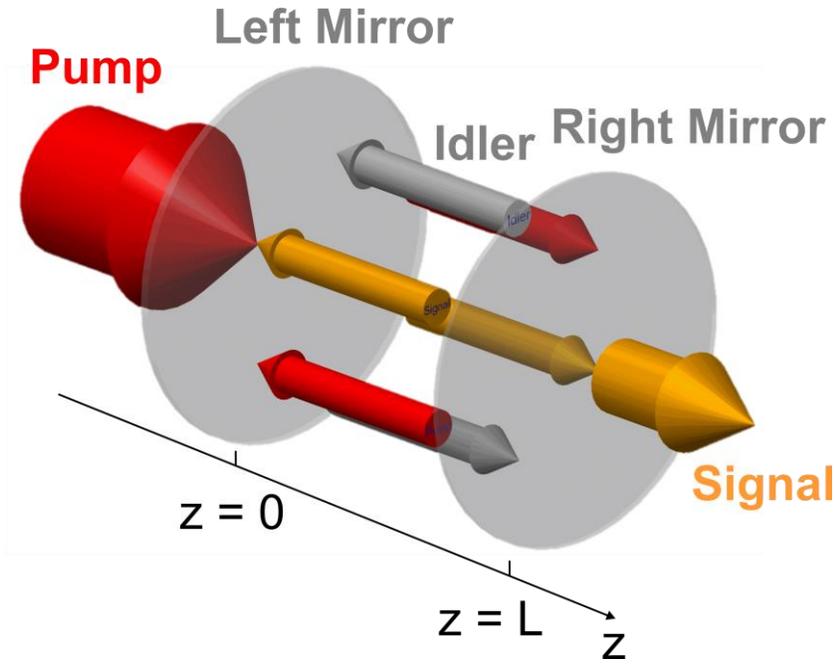

**Figure 8.** Schematic diagram of laser cavity operating using cascade lasing with identified forward and backward propagating pump, signal and idler waves.

The pump, idler and signal waves are trapped between the cavity mirrors and interact with the dopant ions (Fig.8). If the fiber is aligned with the z axis of a rectangular coordinate system the equations that describe the spatial distribution of the pump, signal and idler wave power are given by the following set of 6 ordinary differential equations:

$$\frac{dP_p^\pm}{dz} = \mp\Gamma_p\left[\sigma_{pa}N_1 - \sigma_{pe}N_3\right]P_p^\pm \mp \alpha\,P_p^\pm,$$

(3)

$$\frac{dP(\lambda_1)^\pm}{dz} = \mp\Gamma_{\lambda 1}\left[\sigma_{32a}N_2 - \sigma_{32e}N_3\right]P(\lambda_1)^\pm \mp \alpha\,P(\lambda_1)^\pm,$$

(4)

$$\frac{dP(\lambda_2)^{\pm}}{dz} = \mp\Gamma_{\lambda 1}\left[\sigma_{21a}N_1 - \sigma_{21e}N_2\right]P(\lambda_2)^{\pm} \mp \alpha\ P(\lambda_2)^{\pm}, \tag{5}$$

where '+' and '-' indexes refer to forward and backward propagating waves, respectively. The total power need for evaluating equation (1) and (2) are obtained by adding the power longitudinal power distributions for the forward and backward propagating waves. The boundary conditions for equations (1) –(5) and cavity shown in Fig.8 are given by:

$$\begin{aligned} P_p^+(z=0) &= r_p(z=0)P_p^-(z=0) \\ P_p^-(z=L) &= r_p(z=L)P_p^+(z=L) \\ P^+(\lambda=\lambda_1, z=0) &= r_{\lambda 1}(z=0)P^-(\lambda=\lambda_1, z=0) \\ P^-(\lambda=\lambda_1, z=L) &= r_{\lambda 1}(z=L)P^+(\lambda=\lambda_1, z=L) \\ P^+(\lambda=\lambda_2, z=0) &= r_{\lambda 2}(z=0)P^-(\lambda=\lambda_2, z=0) \\ P^-(\lambda=\lambda_2, z=L) &= r_{\lambda 2}(z=L)P^+(\lambda=\lambda_2, z=L) \end{aligned} \tag{6}$$

Several techniques have been proposed in the literature for the solution of the equations (1) – (6) [33]. One of them is referred to as a coupled solution method (CSM) and has been adapted from the field of high power laser analysis [34,35]. CSM solves equations (3) - (5) by setting non-zero guess values to powers of forward propagating waves at z = 0, whilst zeroing the backward propagating wave powers. In the next step the equations (3) - (5) are integrated for forward propagating waves only from z = 0 to z = L, using a standard algorithm for the solution of a set of ordinary differential equations, e.g. a Runge-Kutta method, thus solving a standard initial value problem (Fig.8, Fig.9). At the right mirror the boundary conditions (6) are applied and the values of backward propagating wave powers at z = L are calculated from the values of forward propagating wave powers at z = L. These values are then used as initial values for integration of the equations (3) - (5) for backward propagating waves from z = L to z = 0 (Fig.8, Fig.10). At the left mirror the boundary conditions (6) are applied and the values of forward propagating wave powers at z = 0 are calculated from the values of backward propagating wave powers at z = 0. This gives a new update of the forward propagating waves powers at z = 0. These values can be used to calculate a residual based on a difference between the newly updated values and those initially guessed. If the residual is larger than the predefined threshold another iteration of the method is carried out by integrating forward waves to the right mirror, applying boundary condition, propagating the backward waves to the left mirror and applying the boundary conditions. After completing another round trip the residual check may be performed again. The iterations of CSM are continued until the prescribed residual is reached. A method that is in principle similar to CSM is a relaxation method (RM).

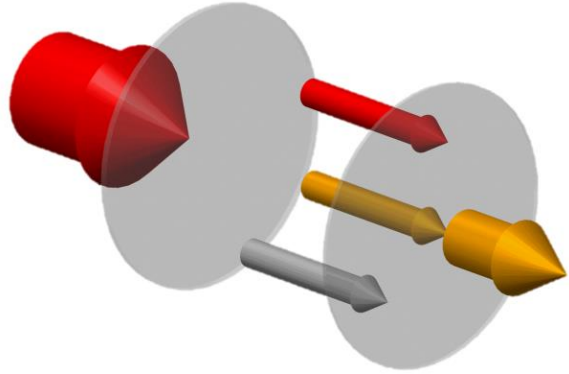

**Figure 9.** Schematic diagram illustrating the principle of coupled solution method of cascade lasing for forward propagation.

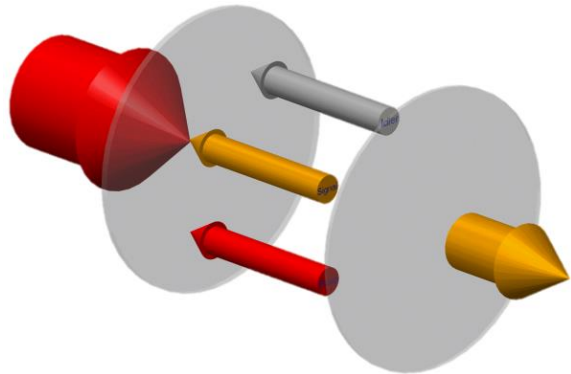

**Figure 10.** Schematic diagram illustrating the principle of coupled solution method of cascade lasing for backward propagation.

In RM the algorithm is initiated by setting non-zero guess values to powers of all waves at z = 0. One can use the boundary conditions (6) to do so consistently. In the next step the equations (3) - (5) are integrated for all waves from z = 0 to z = L (Fig.8, Fig.11). At the right mirror the boundary conditions (6) are applied and the values of backward propagating wave powers at z = L are calculated from the values of forward propagating wave powers at z = L. These values are then used as initial values for integration of the equations (3) - (5) for all waves from z = L to z = 0 (Fig.8, Fig.11). At the left mirror the boundary conditions (6) are applied and the values of forward propagating wave powers at z = 0 are calculated from the values of backward propagating waves powers at z = 0. This gives a new update of the forward propagating waves powers at z = 0 and a possibility to calculate a residual. Again the RM iterations are continued until the prescribed residual is reached.

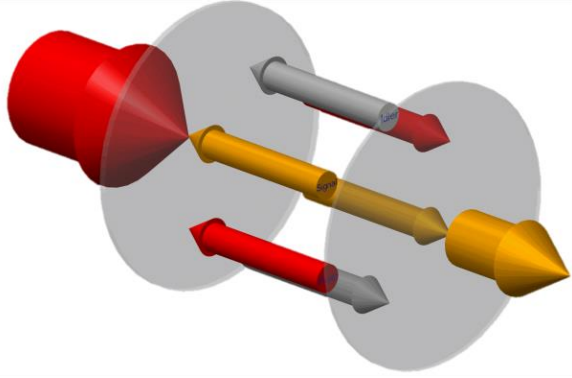

**Figure 11.** Schematic diagram illustrating the principle of relaxation method when applied to analysis of cascade lasing.

The third method that has been used for analysis of chalcogenide glass fiber lasers is the Shooting Method combined with Newton-Raphson Method (SM-NRM). SM-NRM relies on the observation that the integration of equations (3)-(5) from z = 0 to z = L sets an implicit functional relationship between the values of the powers at z = 0 with those at z = L. In order to obtain a set of implicitly formulated nonlinear algebraic equation one only needs to apply boundary conditions (6). This yields a set of implicitly formulated nonlinear algebraic equations whereby the unknowns are the power values for all waves at z = 0. The value of the nonlinear functions is calculated by integrating all powers from z = 0 to z = L according to (3) –(5) and applying (6) at z = L. Since the nonlinear functions are not given explicitly the Jacobian matrix elements have to be calculated using the finite difference method. Similarly, as in the case of CSM and RM the subsequent iterations of Newton-Raphson method are continued until the desired value of the residual is reached.

Figure 12 shows results of a comparative study for all three methods. The CSM and RM have similar properties. They show a good initial convergence both for the signal and the idler. However, after several initial iterations the SM-NRM converges much faster than CSM and RM. We note that in the case of the idler wave a fast convergence rate of SM-NRM results in approaching after 9 iterations the numerical noise floor for double precision arithmetic, which was used in simulations, and hence an apparent growth of the absolute error in the iteration 10 is observed. The comparison of the CPU time shows that SM-NRM is more costly per iteration step than CSM and RM. Further, study also reveals that CSM and RM are less sensitive than SM-NRM to the quality of the initial guess. Hence, an optimal strategy is to combine either CSM or RM with SM-NRM. In such algorithm either CSM or RM is used to perform several initial iterations and the algorithm is switched to SM-NRM. The main characteristics of SM-NRM, RM and CSM are summarized in Table 1.

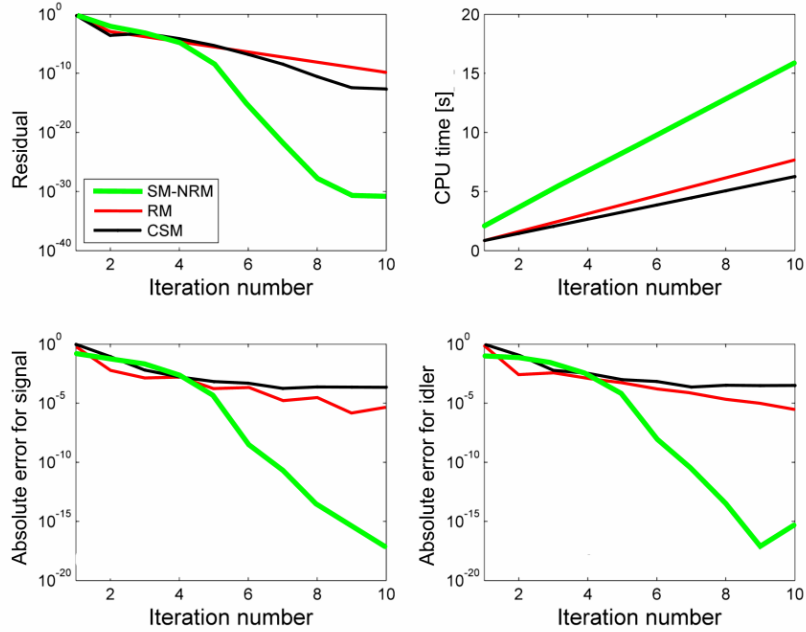

**Figure 12.** Numerically calculated dependence of the residual, CPU time and absolute error for the three algorithms studied: CSM, RM and SM-NRM. The pump power equals 5 W. Figure adopted from [33].

**Table 1.** A summary of main characteristics of SM-NRM. RM and CSM

| parameter | SM-NRM | RM | CSM |
|---|---|---|---|
| Initial rate of convergence | low | moderate | moderate |
| Rate of convergence near solution | high | moderate | moderate |
| Sensitivity to initial guess | high | low | low |
| Amount of CPU time per iteration | moderate | low | low |

## 4. Summary

A review of cavity designs and numerical modelling techniques for lanthanide ion doped chalcogenide glass fiber lasers was presented. The comparative study of several algorithms shows that the best numerical efficiency is achieved by combining Shooting Method - Newton-Raphson Method with either Relaxation Method or Coupled Solution Method. The currently available cavity designs predict the slope efficiency in the range of between 30 % and 40 % for chalcogenide fibers with experimentally demonstrated doping levels. According to the results obtained, output powers in the range of several hundreds of mW should be realizable. However, the experimental efforts so far have not been successful. Among the possible reasons for the lack of success might be the intrinsic fiber loss, which might be higher at the laser operating wavelength than what one could expect estimating the fiber loss at an adjacent wavelength. This is due to presence of impurities, like SeH. Further, the thermal properties of the fibers and their ability to handle high power density is still a subject of intensive research. It is possible that the further improvement of fiber fabrication technology with an increased knowledge of thermal and optical properties will lead, in near future, to the realization of the first lanthanide ion doped chalcogenide glass MIR fiber laser.

**Acknowledgments:** The author acknowledges support from the European Union's Horizon 2020 research and innovation programme under the Marie Skłodowska-Curie grant agreement No. 665778 (National Science Centre, Poland, Polonez Fellowship 2016/21/P/ST7/03666).

**Conflicts of interests:** The author declares no conflict of interests